\documentclass[twocolumn]{aastex62}

\received{\today}
\submitjournal{ApJS}

\shorttitle{Solar Bayesian Analysis Toolkit}
\shortauthors{Anfinogentov et al.}

\usepackage{amsmath}
\usepackage{listings}
\begin{document}

\title{Solar Bayesian Analysis Toolkit --- a new Markov chain Monte Carlo IDL code for Bayesian parameter inference}

\correspondingauthor{Sergey~A.  Anfinogentov}
\email{anfinogentov@iszf.irk.ru}

\author[0000-0002-0786-7307]{Sergey A. Anfinogentov}
\affil{Institute of Solar-Terrestrial Physics SB RAS, Lermontov St. 126, Irkutsk 664033, Russia}

\author[0000-0001-6423-8286]{Valery~M. Nakariakov}
\affil{Centre for Fusion, Space and Astrophysics, Physics Department, University of Warwick, Coventry CV4 7AL, UK}
\affil{St Petersburg Branch, Special Astrophysical Observatory, Russian Academy of Sciences, St  Petersburg, 196140, Russia}

\author[0000-0002-0338-3962]{D. J. Pascoe}
\affil{Centre for mathematical Plasma Astrophysics, Mathematics Department, KU Leuven, Celestijnenlaan 200B bus 2400, B-3001 Leuven, Belgium}

\author{Christopher R. Goddard}
\affil{Germany}



\begin{abstract}
We present  the Solar Bayesian Analysis Toolkit (SoBAT) which is a new easy to use tool for Bayesian analysis of observational data, including parameter inference and model comparison.
SoBAT  is aimed (but not limited) to be used for the analysis of solar observational data.
We describe a new Interactive Data Language (IDL) code designed to facilitate the comparison of user-supplied model with data.
Bayesian inference allows prior information to be taken into account.
The use of Markov chain Monte Carlo (MCMC) sampling allows efficient exploration of large parameter spaces and provides reliable estimation of model parameters and their uncertainties.
The Bayesian evidence for different models can be used for quantitative comparison.
The code is tested to demonstrate its ability to accurately recover a variety of parameter probability distributions.
Its application to practical problems is demonstrated using studies of the structure and oscillation of coronal loops.
\end{abstract}



\section{Introduction}
\label{sec:intro}

The use of Bayesian analysis and Markov chain Monte Carlo (MCMC) sampling is increasingly common in astronomy \citep[e.g. review by][]{2017ARA&A..55..213S}  and heliosesmology \citep[e.g. ][]{2010MNRAS.406..767B,2015MNRAS.454.4120H}.
However, it is not widely used in other branches of solar physics, with exception of magnetohydrodynamic (MHD) seismology of the solar corona, where the advantages of the Bayesian approach are intensively exploited.
The details can be found in a recent review  considering the use of Bayesian analysis for coronal seismology in particular \citep{2018AdSpR..61..655A}.

Traditionally, the problem of estimating model parameters from observational data (parameter inference) is solved by the best fitting approach which aims to find in the parameter space a  point giving the best agreement between the model and observations.
This is usually done by computing the maximum likelihood estimate (MLE) or least squares estimate (LSE) which is  equal to MLE  in the case of the normally distributed measurement errors.
Thus, the aim of the best fitting approach is to find in the parameter space the global maximum corresponding to the best fit of the model  to the observed data.
The Bayesian approach is different: instead of searching for the highest peak in the parameter space, it implies making a map of the whole parameter space in the form of posterior probability distribution function (PDF) representing all  information available from both observations and prior knowledge.
This function gives a probability density for every point in the parameter space reaching a global maximum at the position corresponding to the best fitting combination of model parameters.

This lead us to  the main advantage of the Bayesian approach which is a correct estimation of the uncertainties.
Although, least squares fitting software often provides uncertainties estimation based on some assumptions like  the Gaussian shape of a parameter distribution, such an estimation became incorrect  when these assumptions are not valid, for example, if the parameter distribution significantly differs from the normal one (e.g. asymmetric or multi-modal).
Since the Bayesian analysis is capable to recover even a complex parameter distribution being very different from the normal one,  it allows for correct and reliable estimation of the uncertainties for a broad range of parameter inference problems.

Often, there are more than one models that can explain observational data.
In this case, one needs to have a possibility to quantitatively  compare competing models.
A good model should have the following properties:
\begin{enumerate}
\item The best fit produced by the model should be close to the observed data points.
\item The model should not be over-fitted by having too many free parameters.
\item It should be confined in the parameter space. The model parameters should be well constrained based on the observational data.
\item It should be confined in the observational data space. The model should not predict observations far away from the actual data points.
\end{enumerate}
To assess a model within the traditional best fitting approach   the reduced $\chi^2$  criterion is mainly used.
Though it allows us to assess the best fits (point~1) and accounts for the number of model parameters (point~2), it does not take into account the last two items from the list above and ignores the model confinement in the parameter and data spaces.
Opposite to this, the Bayesian analysis offer a model comparison criterion called Bayes factor that assesses the whole models but not only the best fits and transparently accounts for all four properties mentioned in the list above. 

The advantage of the Bayesian approach could be illustrated by the following specific example.
In coronal seismology, one of the standard operations is the determination of parameters of kink oscillations.
Suppose the observations gives us a time series of the oscillating displacements of a coronal loop.
Theory predicts that the oscillation could be damped by either exponential or Gaussian law, and that the oscillation could be a superposition of several harmonics.
Thus, the observationally obtained time series could be approximated by several different theoretically prescribed functions.
For each specific function, its parameters that best fit the data could be determined by the MLE or LSE. However, the Bayesian analysis allows us to compare the quality of fittings by those different functions with each other. 

The aim of this work is to provide the solar physics community with a reliable and easy to use tool for Bayesian analysis of observational data, including parameter inference and model comparison. 
Although, there are few efforts to bring Bayesian methodology to the IDL community (see e.g. idl\_emcee sampler at \url{https://github.com/mcfit/idl_emcee}), according to our knowledge our IDL code provides unique features such as high level routines for \lq\lq fitting\rq\rq\ observational data and numerical tools for Bayesian model comparison.

This paper is organised as follows; the Bayesian method and techniques used in the code are presented in Sect.~\ref{sect:method}.
Tests of the sampling algorithm are performed in Sect.~\ref{sect:tests}.
The code is demonstrated by applying it to simple test problems in Sect.~\ref{sect:examples}, and to practical solar physics problems in Sect.~\ref{sect:applications}.
Concluding remarks are presented in Sect.~\ref{sect:conclusions}.

\section{Bayesian approach to parameter inference}
\label{sect:method}

A parameter inference problem implies that the observed data $D$ can be explained in terms of the model $M$ (i.e. an analytical function such as a sinusoid, a Gaussian, or even an underlying numerical code) having a parameter set $\theta = [\theta_1, \theta_2, ..., \theta_n]$.
For example, in the case of a sinusoidal function, $\theta_i$ can be the values of the period, amplitude, and phase.
Thus, the aim is to find the value of the parameters $\theta$ that gives the best possible agreement with the observed data $D$.
The formulation of the Bayesian parameter inference relies on three main definitions:
\begin{enumerate}
\item \textit{The prior probability density function (PDF)} $P(\theta)$ represents our knowledge about the model parameters $\theta$ before considering the observational data $D$.
For example, this could be knowledge from previous measurements or a requirement that the particular model parameter lies inside a certain range.

\item \textit{The sampling PDF} $P(D|\theta)$ describes the conditional probability to obtain the observed data $D$ given that the model parameters $\theta$ are fixed.
The  sampling PDF is closely related to the measurement errors.
For example, if measurement errors in our experiment follow (or can be assumed to follow) the normal distribution, the sampling PDF would be a normalised Gaussian.  

\item \textit{The likelihood function} is literately  the sampling PDF  $P(D|\theta)$ considered as a function of $\theta$ with fixed $D$.
We note that in contrast to the sampling PDF, the likelihood function is not a probability density.
In particular, its integral over $\theta$ is not equal to unity. To become a posterior PDF, the likelihood function needs to be normalised.

\item \textit{The posterior PDF} $P(\theta|D)$ describes the conditional probability that the model parameters are equal to $\theta$ under condition of observed data being equal to $D$.
This function represents our knowledge on the model parameters $\theta$ after the observation, when the observed data $D$ is known and fixed.
\end{enumerate}

The Bayes theorem connects prior and posterior probability density functions and describes how the observational data $D$ affects our knowledge about model parameters $\theta$
\begin{equation} \label{eq:Bayes theorem}
P(\theta|D) = \frac{P(D|\theta)P(\theta)}{P(D)}.
\end{equation}

The normalisation constant $P(D)$ is the \textit{Bayesian Evidence} or marginalised likelihood
\begin{equation} \label{eq:Evidence}
P(D)=\int P(D|\theta)P(\theta) d\theta
\end{equation}
For our prescribed prior probability $P(\theta)$ and likelihood $P(D|\theta)$ functions, the posterior probability distribution $P (\theta|D)$ can be readily computed for any value of the parameter set $\theta$ using the Bayes theorem in Eq. (\ref{eq:Bayes theorem}).
However, in practical applications, we are interested in finding an estimate and corresponding uncertainties for each parameter $\theta_i$.

The most common choice in Bayesian statistics for an estimate of unknown parameters $\theta$ is a maximum a posteriori probability (MAP) estimate $\theta_{MAP}$ which is a point in the parameter space where the posterior PDF reaches its global maximum.
Other estimates e.g. the expected value or the median can be also used.

To put uncertainties around the estimate, one needs to calculate the marginalised (integrated) posteriors
\begin{equation} \label{eq:Marginalised posterior}
P(\theta_i|D) =\int P(\theta_1,\theta_2, ... ,\theta_N|D)d\theta_{k\neq i}.
\end{equation}
For a simple low-parametric model (2--3 parameters), the multiple integrals in Eq. (\ref{eq:Marginalised posterior}) can be directly calculated using standard numerical methods.
Unfortunately, it is practically impossible to use direct numerical integration for complicated models with a large set of parameters.
Indeed, every additional parameter increases the computation time by several orders of magnitude.
Therefore, sampling methods based on MCMC are preferable for complex models. MCMC allows us to obtain samples from the posterior probability distribution $P(\theta|D)$.
When enough samples are obtained, the marginalised posterior (Eq. (\ref{eq:Marginalised posterior})) can be approximated by a histogram of the corresponding model parameter $\theta_i$.

\subsection{Posterior Prediction}
\label{sec: Posterior Prediction}

Once the most credible value $\theta_{\rm MAP}$ of the model parameters is determined, one can calculate the predictive distribution of observational data points (i.e. what the next observation $D_{\rm new}$ could be):
\begin{equation}\label{eq:predictive distribution}
P(D_{\rm new}|\theta_{\rm MAP}) = P(D_{\rm new}|\theta = \theta_{\rm MAP}).
\end{equation}
However, Equation (\ref{eq:predictive distribution}) does not account for the estimate $\theta_{\rm MAP}$ being uncertain itself.
This uncertainty comes from the observational errors and model limitations, and is the width of the Posterior PDF in the vicinity of its global maximum.
To account for all uncertainties correctly, the Posterior Predictive Distribution
\begin{equation}\label{eq:posterior_predictive_distribution}
P(D_{\rm new}|D) = \int P(D_{\rm new}|\theta) P(\theta|D)d\theta 
\end{equation}
is used. It is usually broader than the distribution given by Equation (\ref{eq:predictive distribution}) because of the  additional uncertainties in $\theta$.

The Posterior Predictive Distribution can be used for two purposes.
First one is to forecast future observations and to provide reliable prediction intervals, if the model allows for extrapolation in time.
The second application is a so called Posterior Predictive check, which allows for assessing the consistency of the chosen model with the observations in terms of confinement of the model in the data space.
A reliable model should produce a narrow distribution predicting possible observations of the same process to be close to the actual data points.

\subsection{Model comparison}
\label{sec:model_comparison}

Bayesian analysis allows for quantitative comparison of two models $M_1$ and  $M_2$ by calculating the \textit{Bayes factor} \citep{Jeffreys1961}, defined as
\begin{equation} \label{eq:Bayes factor}
B_{12} = \frac{P(D|M_1)}{P(D|M_2)},
\end{equation}
where the evidences $P(D|M_1)$ and  $P(D|M_1)$ are calculated according to Eq. \ref{eq:Evidence}.
Traditionally, the doubled natural logarithm of this factor is used, i.e.
\begin{equation}
K_{12} = 2\ln B_{12},
\end{equation}
where  values  of $K_{12}$ greater  than  2,  6,  and  10  correspond  to \lq\lq positive\rq\rq , \lq\lq strong\rq\rq, and \lq\lq very strong\rq\rq\  evidence for model $M_1$ over model $M_2$, respectively \citep{Kass1995}.

\section{Description of the code}

SoBAT consists of the following subroutines and functions;

\begin{itemize}

\item \texttt{MCMC\_FIT} is a high-level routine used to fit $y = f(x, \theta)$ dependence to the measured data points $[X_i, Y_i]$  with normally distributed measurement errors $N(0, \sigma)$. The errors $\sigma$ can either be provided by the user via the \texttt{ERRORS} keyword or automatically inferred as an additional free parameter.
The input parameters are the observational data points $[X_i, Y_i]$, initial guesses for the free parameters of the model, the IDL function implementing $y = f(x)$ dependence, and an array of priors for each parameter.
The generated samples will be returned in the \texttt{SAMPLE} keyword parameter.

\item \texttt{MCMC\_FIT\_EVIDENCE} function can be used to calculate the Bayesian evidence (\ref{eq:Evidence}) from the output of the  \texttt{MCMC\_FIT} subroutine. The input parameters are generated samples, data points $[X_i, Y_i]$, priors and an IDL function implementing $y = f(x)$ dependence.
The function returns the calculated evidence as a scalar value.

\item \texttt{MCMC\_SAMPLE} is a low level function which generates samples from a  target function provided by the user.
This function allows the user to sample a custom posterior PDF and should be used for the cases where the observed data can not be modelled as  $y = f(x,\theta) + N(0,\sigma)$. 
The input parameters of \texttt{MCMC\_SAMPLE} function are the initial guess, the IDL function that calculates target PDF to sample, and the number of samples to generate.
The \texttt{MCMC\_SAMPLE} returns generated samples as an array.

\item \texttt{MCMC\_EVIDENCE} function can be used to calculate the Bayesian evidence (Eq.~\ref{eq:Evidence}) from the output of the  \texttt{MCMC\_FIT} subroutine. The input parameters are the IDL function calculating the posterior PDF and samples array returned by the \texttt{MCMC\_SAMPLE} function. The computed evidence is returned as a scalar number.

\item Functions for constructing priors, namely \texttt{PRIOR\_UNIFORM}, \texttt{PRIOR\_NORMAL}, \texttt{PRIOR\_HALFNORMAL}, and \texttt{PRIOR\_EXPONENTIAL}, allow to setup prior distributions for the free parameters.
SoBAT also provides the \texttt{PRIOR\_CUSTOM} routine, which allows to pass a user defined IDL function as a prior PDF.

\end{itemize}

\subsection{Sampling algorithm}
\label{sec:sampling}

To generate a large number of samples from the posterior distribution, SoBAT uses the Markov chain Monte Carlo technique.
The marginalised posterior PDFs are than approximated by the histograms of these samples.

The MCMC sampling algorithm is the most important part of our code.
It can generate samples from the posterior distribution using any target function $f(\theta)$ which is proportional to the posterior PDF $P(\theta|D)$  and is a known continuous function that can be calculated for any value of $\theta$. Thus, the knowledge of the normalisation constant (Eq.~\ref{eq:Evidence}) is not required for the inference. 

Our sampling algorithm  is the classical random walk Metropolis-Hasting  sampler with the multivariate normal distribution  used as a proposal  distribution. 
Its covariance matrix  $\hat{\sigma}$ is automatically tuned to keep the acceptance rate in the range of 10 -- 90\,\% during the whole sampling procedure. 
In order to generate the whole sequence of samples (chain) with the same proposal distribution, we restart the sampling procedure every time when the proposal distribution is tuned.
The detailed description of the algorithm is given below:

\begin{enumerate}

\item Initialise the starting point in the parameter space, $\Theta_0$.

\item Estimate the local covariance matrix $\hat{\sigma}$ for $\theta = \Theta_0$.

\item \label{lst:rpt_start} Simulate the proposed sample $\Xi_i$ from the multivariate normal distribution $N(\Theta_i, \hat{\sigma})$ with the expected value $\Theta_i$ and covariance matrix $\hat{\sigma}$.

\item Compute the ratio $R = f(\Xi_i|D) / f(\Theta_i|D)$.

\item Peak a random number $\varepsilon$ between 0 and 1.

\item Produce a new sample $\Theta_{i+1}$:$$
        \begin{cases} 
        \mathrm{accept:}\ \Theta_{i+1} = \Xi_i;\  N_\mathrm{a} = N_\mathrm{a} +1; &(\mathrm{if}\ \varepsilon \le  R  )\\
         \mathrm{reject:}\ \Theta_{i+1} = \Theta_i;\ N_\mathrm{r} =N_\mathrm{r} +1; &(\mathrm{if}\ \varepsilon >  R) .
        \end{cases}
        $$

\item Calculate the acceptance rate $r = N_\mathrm{a}/( N_\mathrm{r}+ N_\mathrm{a})$.

\item \label{lst:rpt_stop}  if $r < 10\,\%$ or  $r > 90\,\%$\footnote{For a particular problem this range can be tuned.} then set $\Theta_0 = \Theta_{i+1}$ and  go to step 2.

\item Repeat steps \ref{lst:rpt_start}--\ref{lst:rpt_stop} until the desired number of samples is generated.

\item Return all collected samples $\Theta_i$ as a result.

\end{enumerate}

After several restarts, the sampling algorithm usually finds the maximum probability area and stabilise there with acceptance rate about 10\% -- 90\%. 
We should note, that there is no guaranty that the algorithm will find the global maximum for a given number of iterations.
Therefore, we recommend providing a rather good initial guess and to generate a sufficiently large number of samples.

\subsubsection{Burning in stage}

The developed code runs the sampling procedure twice.
The first run is so called \lq\lq burning in\rq\rq\ and is used to allow the chain to explore the parameter space and  to converge to the global probability maximum in the parameter space.
The second chain (main sampling) starts from the high probability area found during the burning in stage and may use the  samples obtained during the first run to construct the optimal proposal distribution. 
The chain collected during the main sampling is then returned as a sampling result.

\subsection{Estimation of the proposal distribution}

The selection of the proposal distribution is essential for constructing an effective Metropolis-Hastings sampler.
The developed code uses the multivariate normal distribution with the expected value $\mu = \Theta_0$ and the covariance matrix $\hat{\sigma}$, which is tuned to reflect the local properties of  the parameter space and to achieve an optimal acceptance rate. 
The algorithm of the calculation of the optimal covariance matrix  $\hat{\sigma}$ is given below.

\begin{enumerate}
	\item Initialise variables.
	\begin{itemize}
		\item $\Theta_0$ -- a position in the parameter space 
		\item $\hat{\sigma}$ -- an initial guess for the covariance matrix
		\item $S$ -- an array to store generated samples 
	\end{itemize}
	\item Simulate the proposed sample $\Xi_i$ from the multivariate normal distribution $N(\Theta_0, \hat{\sigma})$ with the expected value $\Theta_0$ and covariance matrix $\hat{\sigma}$.
		\item Compute the ratio $R = \min{\left(\frac{f(\Xi_i|D)}{f(\Theta_0|D)},\frac{f(\Theta_0|D)}{f(\Xi_i|D)}\right)}$.
		\item Generate a random number $\varepsilon$ between 0 and 1.
		\item If $\varepsilon \le  R$, accept and save  sample $S\leftarrow \Xi_i$; $N_\mathrm{a} = N_\mathrm{a}+1$ or reject it $N_\mathrm{r} = N_\mathrm{r}+1$  otherwise.
		\item Calculate the acceptance rate $r= N_\mathrm{a}/( N_\mathrm{r}+ N_\mathrm{a})$.
		\item Tune $\hat{\sigma}$ for better acceptance rate
		\begin{itemize}
			\item if $r = 0$ during 500 subsequent iterations, set $ \hat{\sigma} =  0.5 \hat{\sigma}$
			\item if  $r > 50\%$,  set  $ \hat{\sigma} =  1.1 \hat{\sigma}$
		\end{itemize}
		\item If more than 500  samples were accepted, set $\hat{\sigma} = \mathrm{covariance}(S)$ 			
    \item Repeat steps 2--8 until the desired number of samples is generated.
	\item Return $\mathrm{covariance}(S)$ as a result.
\end{enumerate}

\subsection{Quantitative model comparison}

The code allows evidences to be calculated by numerical evaluation of the integral given by Eq.~(\ref{eq:Evidence}).
The ratio of evidences for two models is the Bayes factor and can be interpreted as described in Sect.~\ref{sec:model_comparison}.
The numerical integration of Eq.~(\ref{eq:Evidence}) is implemented using the importance sampling Monte-Carlo technique \citep{10.1093/biomet/57.1.97}.
As an importance function, we use a multivariate Gaussian  with the covariance matrix computed from the simulated MCMC samples from the posterior distribution.

To compute evidence for a given model, SoBAT offers the \texttt{MCMC\_EVIDENCE} function. 
The function has three required parameters:
\begin{itemize}
\item  $f(\theta)$ -- a function computing the natural logarithm of a target function  proportional to the posterior PDF;
\item  $S_i,  i=1..N_s$ -- Samples simulated from the posterior by the \texttt{MCMC\_SAMPLE} function;
\item $N$ -- Number of iteration for the Monte-Carlo integration.
\end{itemize}
The importance sampling Monte-Carlo integration is interpreted in the following form:
\begin{enumerate}
\item Estimate the covariance matrix $[\hat{\sigma}]$ and the expected value $[\mu]$ from the posterior samples. The PDF $n(\theta)$ of the  the multivariate normal distribution $N(\mu, \hat{\sigma})$ will be used as the importance function. 
\item Repeat N times ($i = 1..N$):
\begin{enumerate}
\item Simulate a position\footnote{Here $\theta_i$ denotes the full vector of free parameters}  $\theta_i$ in the parameter space from the multivariate normal distribution $N(\mu, \hat{\sigma})$;
\item Compute the  value of the importance function for the current position $g_i = n(\theta_i)$;
\item Compute the  target function $f(\theta)$ for the current position in the parameter space $f_i=f(\theta_i)$.
\end{enumerate}
\item The integration result is calculated as $\frac{1}{N} \sum_{i=1}^N \frac{f_i}{g_i}$.
\end{enumerate}

Here, importance sampling is used to improve the convergence of the Monte-Carlo integration. The form of the specific importance function does not have any implication for the posterior PDF.
Therefore, though we use the multivariate Gaussian as the importance function, the posterior PDF can still be an arbitrary function more or less confined in the parameter space.

\subsection{Fitting functions}

  One of the most frequent application of the Bayesian analysis and MCMC is to infer parameters $\theta$ of a model $M$ which is an analytical function that describes theoretical dependence of $y$ upon $x$ and has a set of free parameters $\theta$:
  $$
  y = M(x, \theta)
  $$

  from the observed data points ($D=[X_i,Y_i] : i=1..N$) where $N$ is the number of data points, $X_i$ and $Y_i$ are empirically determined values of $x$ and $y$ in the $i$-th measurement. The uncertainties of the fitted parameters $\theta = [\theta_1, \theta_2, \cdots,  \theta_{N_p}]$ have also to be estimated.
  SoBAT contains the (\texttt{MCMC\_FIT}) routine which is aimed to solve this problem.
  
  MCMC\_FIT utilises the assumption that the error corresponding to  $Y$ measurements is normally distributed with the standard deviation  $\sigma_Y$.
  Thus, the likelihood function is the product of $N$ Gaussians
  \begin{equation}
  P(D|\theta) = \frac{1}{(2\pi \sigma_Y^2)^{\frac{N}{2}}}\prod_{i=1}^{N} \exp \left\lbrace  -\frac{[Y_i - M(X,\theta)]^2}{2\sigma_Y^2}\right\rbrace.
  \label{eq:mcmc_fit likelihood}
  \end{equation}
  The measurement error $\sigma_Y$ is considered as one of the unknown parameters.
  It is also assumed  to be the same for all data points  and is inferred during the MCMC simulations together with $\theta$.
  
  As an a priori knowledge, a user can provide a  range of the possible model parameter values $\theta $:
  $$\theta_i^{\mathrm{min}} \le \theta_i \le \theta_i^{\mathrm{max}}.$$
  
  Thus, our prior probability distribution can be expressed as follows
  \begin{equation}
  P(\theta) = \prod_{i=1}^N H(\theta_i,\theta_i^{\mathrm{min}},\theta_i^{\mathrm{max}}),
  \end{equation}
  where $H(\theta_i,\theta_i^{\mathrm{min}},\theta_i^{\mathrm{max}})$ is the  PDF of a uniform distribution in the range $[\theta_i^{\mathrm{min}},\theta_i^{\mathrm{max}}]$ which is defined as
  \begin{equation}
  H(\theta_i,\theta_i^{\mathrm{min}},\theta_i^{\mathrm{max}}) = \left\{  \begin{matrix}
  \frac{1}{\theta_i^{\mathrm{max}}-\theta_i^{\mathrm{min}}},& \theta_i^{\mathrm{min}} \le \theta_i \le \theta_i^{\mathrm{max}} \\
  0,& \mathrm{otherwise}\\ 
  \end{matrix}\right.
  \end{equation}

\subsection{Posterior predictive check}

One of the ways to check the correctness of the parameter inference is to estimate the Posterior Predictive Distribution, by sampling from it during  the main sampling procedure.
In the \texttt{MCMC\_FIT} routine,  Eq.~(\ref{eq:mcmc_fit likelihood}) is used to generate a sample from the posterior predictive distribution of the measured data $[Y]$ for every sample from the posterior distribution $[P(\theta|D)]$.
In the case of a user supplied posterior PDF, the user is responsible for simulating samples from the predictive distribution within the user supplied IDL function computing posterior PDF and for returning it in the \texttt{ppd\_sample} keyword.

\section{Tests of the sampling algorithm}
\label{sect:tests}

The designed sampling algorithm (see Sect.~\ref{sec:sampling}) uses a multivariate normal distribution as a proposal.
Therefore, the robustness of sampling procedure should be tested on target distributions that are significantly different from the normal distribution.
In this section, we present such tests for univariate and bivariate target densities.

\subsection{1D target distributions} \label{sec: hist 1D}
\begin{figure*} 
	\includegraphics[width =1.\textwidth]{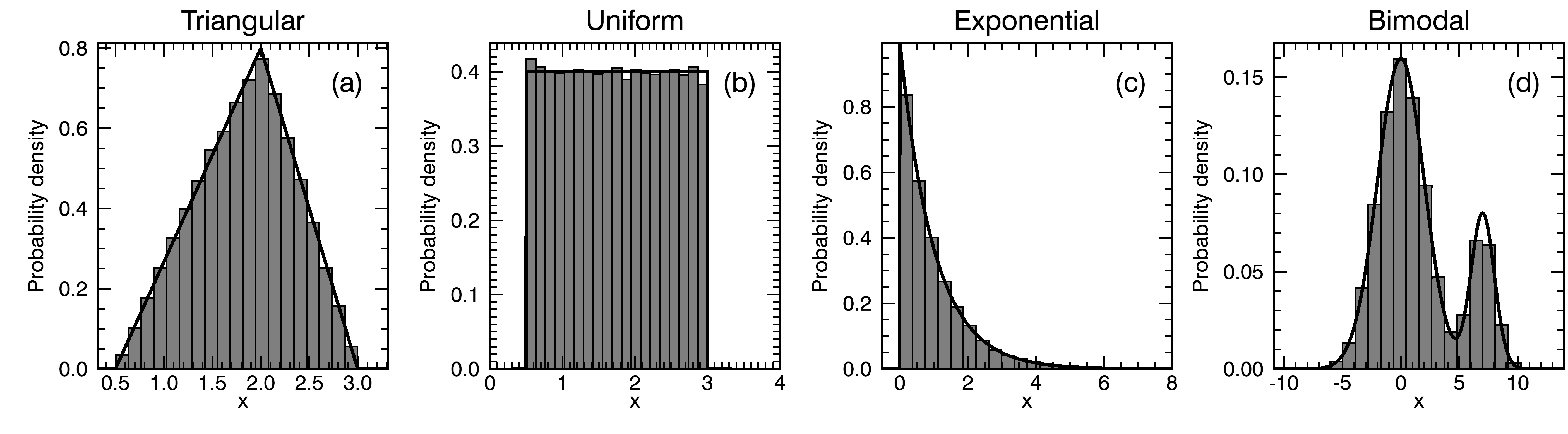}
	\caption{Normalised histograms of $10^5$ MCMC samples obtained from different univariate target distributions: asymmetric triangular (a), uniform (b), exponential (c), and mixture of two normal distributions (d).  The target distributions are plotted over  histograms with solid black lines. }
	\label{fig:hist_1d}
\end{figure*}

To  test the sampling procedure used in the developed code, we selected the following 1D distributions: slightly asymmetrical triangular
$$
f(x) = 
\begin{cases}
\frac{2(x-a)}{(b-a)(c-a)}&\text{for $a < x \leq c$},\\
\frac{2(b-x)}{(b-a)(b-c)} &\text{for $c< x < b$},\\
0 &\text{otherwise}
\end{cases}
$$
  with $a=0.5$, $b = 3$, and $c = 2.5$ (see Fig. \ref{fig:hist_1d}a); uniform
  $$
  f(x) = 
  \begin{cases}
  \frac{1}{(b-a)}&\text{for $a < x < b$},\\
  0 &\text{otherwise}
  \end{cases}
  $$
  with $a=0.5$ and $b=3$ (see Fig. \ref{fig:hist_1d}b);  exponential
  $$
  f(x) = 
  \begin{cases}
  \lambda e^{-\lambda x}&\text{for $x > 0$},\\
  0 &\text{otherwise}
  \end{cases}
  $$
   with $\lambda = 1$ (Fig.~\ref{fig:hist_1d}c); and  a bimodal mixture of 2 normal distributions  with different expected values and dispersions
   $$
   f(x) = 0.8\frac{1}{2\pi\sigma_1^2}e^{\frac{x - \mu_1}{2\sigma_1^2}} + 0.2\frac{1}{2\pi\sigma_2^2}e^{\frac{x - \mu_2}{2\sigma_2^2}}
   $$
 with $\mu_1 =0$,  $\mu_2 = 7$,  $\sigma_1 = 2$ and $\sigma_2 = 1$ (see Fig.~\ref{fig:hist_1d}d).
Normalized histograms of the $10^5$ MCMC samples generated for each distribution are shown in Fig.~\ref{fig:hist_1d}.
The obtained histograms perfectly coincide with the corresponding target densities shown in Fig.~\ref{fig:hist_1d} with solid black lines. 

\subsection{2D target distributions}

\begin{figure*}[ht]
	\includegraphics[width =1.\textwidth]{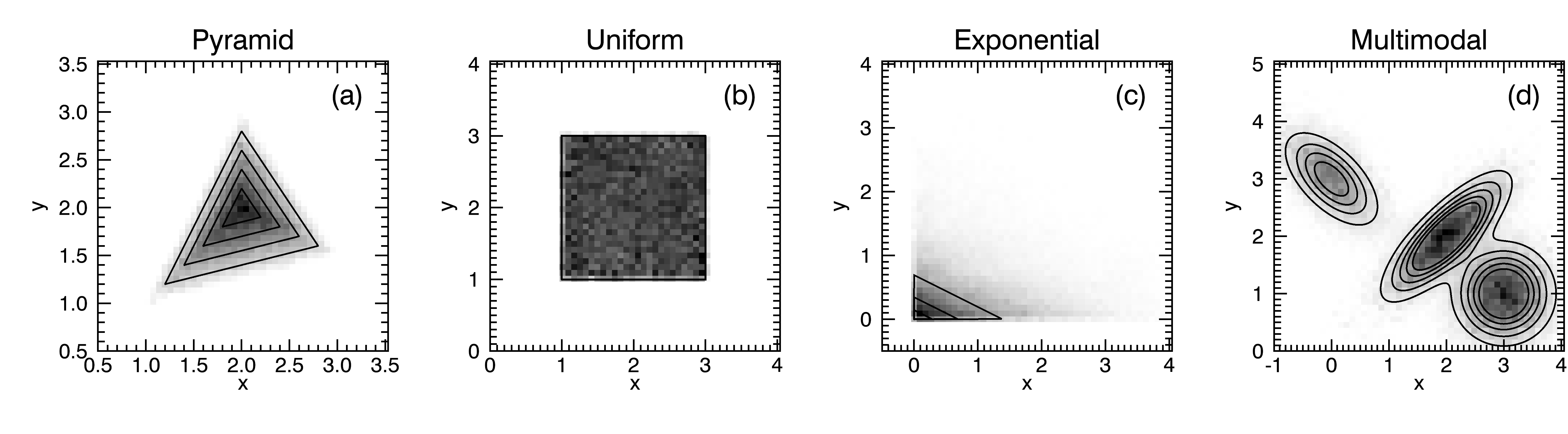}
	\caption{2D histograms (background colour) of $10^5$ MCMC samples obtained from different bivariate target distributions: pyramid (a), uniform (b), exponential (c), and mixture of three normal distributions (d).  The target distributions are shown by contours. }
	 \label{fig:hist_2d}
\end{figure*}

To demonstrate the  correctness of the sampling procedure in multi parametric case, we present the testing results for a set of bivariate target probability densities.
We selected 2D versions of the distributions used in \ref{sec: hist 1D}: pyramid (Fig. \ref{fig:hist_2d}a), 2D uniform distribution bounded by a square (Fig. \ref{fig:hist_2d}a), 2D exponential distribution, and a mixture of 3 bivariate normal distributions with different expected values and covariance matrices.
The 2D histograms (see Fig. \ref{fig:hist_2d}) are perfectly coinciding with the target densities, shown in  Fig. \ref{fig:hist_2d} by contours.

\section{Examples of usage}
\label{sect:examples}

In this section, we demonstrate examples of using SoBAT library to fit a simple linear dependence and consider an example of the Bayesian model comparison.

\subsection{Fitting a linear dependence}
\label{sec:linear_dependence}

Let us consider a simple example of fitting a set of synthetic data points ${X_i, Y_i}$ by a linear function to illustrate the practical usage of SoBAT.
The synthetic data points in our example are generated using the linear dependence with the present of the normally distributed noise
$$Y_i = k X_i + b + N(0,\sigma),$$
where $k =0.5$, $b=1$, and $\sigma = 2$.

Firstly, we need to specify the model as a function describing the linear dependence of $y$ upon $x$.
The model function for the linear dependence is given in Listing \ref{lst:linear_model}.
\begin{lstlisting}[float,caption=Model function for the linear dependence,label={lst:linear_model}]
  function lin_model, x, params
    k = params[0]
    b = params[1]
    return, k * x + b
  end
\end{lstlisting}

\begin{figure}[ht]
 	\includegraphics[width =1.\linewidth]{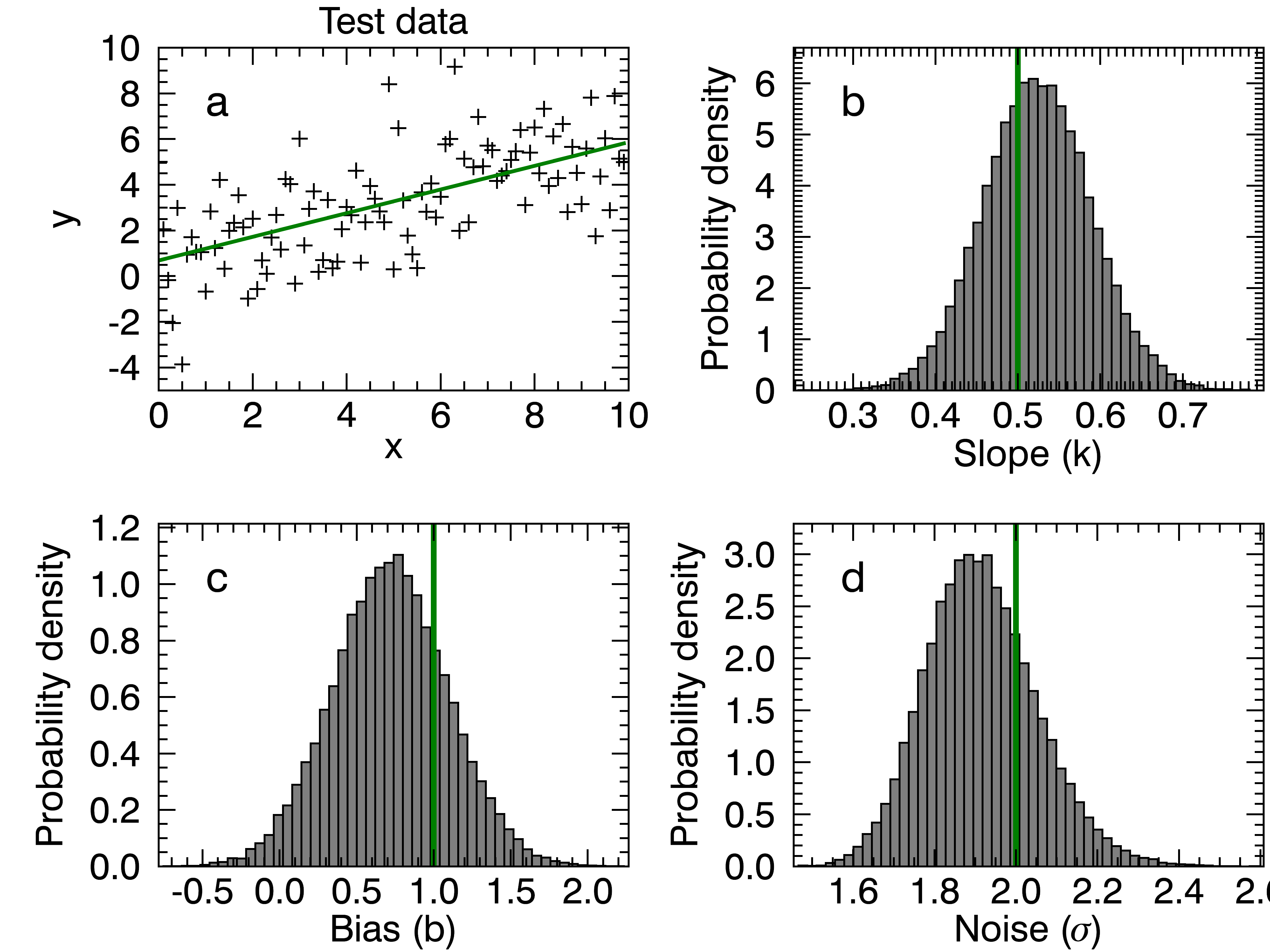}
 	\caption{\textbf{Panel a:} linear dependence $y = k x + b$ (green line) fitted to the noisy synthetic data points (crosses) using the  \texttt{MCMC\_FIT} function.  \textbf{Panels b -- d}: normalised  histograms approximating marginalised posterior distributions of  the  gradient $k$ (b), bias $b$ (c), and noise level $\sigma$ (d) obtained from $10^5$ MCMC samples. True values of the parameters  used to generate synthetic data points are shown by vertical green lines on panels b--d.}
 	\label{fig:hist_linear}
 \end{figure}

Then,  we define allowed limits as uniform priors  and an initial guess for the model parameters $k$, $b$ (lines 2 -- 6 in Listing \ref{lst:linear_mcmc}).
After the call of \texttt{MCMC\_FIT} function (lines 12 -- 14 in Listing \ref{lst:linear_mcmc}), the variable \texttt{fit} will contain the best fitting values for $Y$.
The fitted parameters values and corresponding uncertainties will be stored in the \texttt{pars} and \texttt{credible\_intervals} variables.
The MCMC samples will be returned in  the \texttt{samples} keyword. The latter can be used to plot histograms approximating the marginalised posterior distribution\textbf{s}.
The histograms obtained for the slope ($k$), bias ($b$) and noise level ($\sigma$)  are given  in Figure \ref{fig:hist_linear} (b -- d).
Note, that the true parameter values (green vertical lines in Figure \ref{fig:hist_linear}) do not  coincide with global maximum of the histograms, but lie within the high probability area illustrated by histograms.
Such a behaviour is expected because our inference (as any measurement) is uncertain.
The uncertainty  is described by the width of the histograms and can be quantified for an arbitrary level of significance by computing credible intervals as percentiles of the samples generated with the MCMC code.

\begin{lstlisting}[float,caption=Running MCMC fitting of the linear dependence for the data defined by Listing \ref{lst:linear_model},label={lst:linear_mcmc}]
  ; define priors
  priors = objarr(2)
  priors[0] = prior_uniform(-5d, 5d)
  priors[1] = prior_uniform(-5d, 5d)
  ; define the initial guess
  pars = [1d, 1d]
  ; define the number of samples
  n_samples = 100000
  ; define the number of burn in samples
  burn_in = 10000
  ; run MCMC fitting
  fit = mcmc_fit(x, y, pars, "lin_model", $
    priors=priors, burn_in=burn_in, $
    n_samples=n_samples, samples=samples, $
    credible_intervals=credible_intervals)
\end{lstlisting}

 \begin{figure}[ht]
 \centering
 	\includegraphics[width =0.7\linewidth]{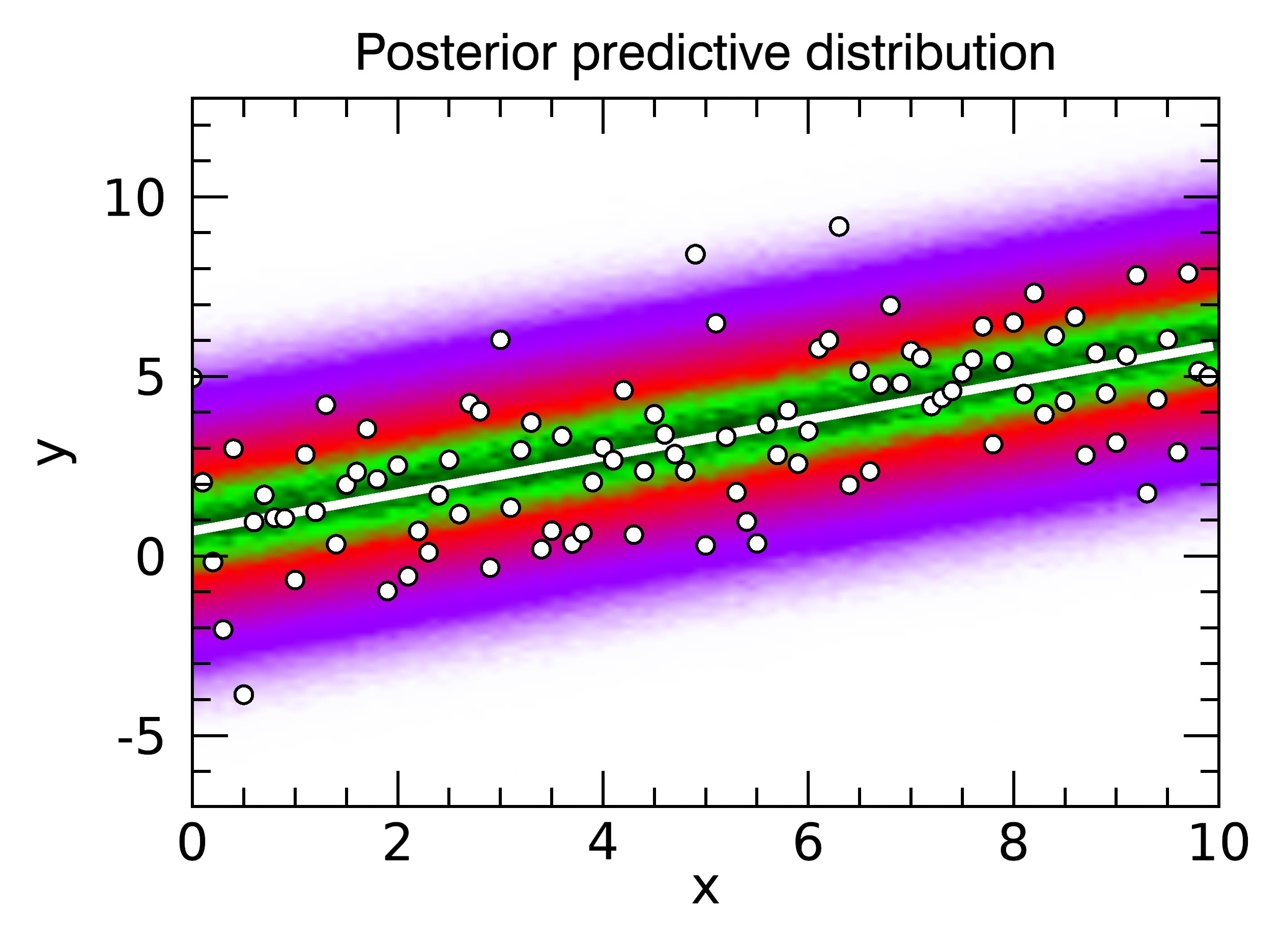}
 	\caption{Posterior predictive probability distribution for a linear dependence fitted to the noisy synthetic data using the \texttt{MCMC\_FIT} function. Data points are indicates by white circles while the white line shows the best fit.}
 	\label{fig:ppd_linear}
 \end{figure}


\subsection{Example of Bayesian model comparison}

To illustrate quantitative comparison of different user-defined models, we use the same synthetic data set as in Sect.~\ref{sec:linear_dependence} with the linear dependence contaminated by white noise. Now we attempt to fit it with a second model with the quadratic dependence:

\begin{equation} 
y = k x + b + c x^2 + N(0,\sigma).
\label{eq:quad_model}
\end{equation}

Listing \ref{lst:quadratic_model} shows the IDL representation of this model.

\begin{lstlisting}[float,caption=Model function for the quadratic dependence,label={lst:quadratic_model}]
  function quad_model, x, params
    k = params[0]
    b = params[1]
    c = params[2]
    return, k * x + b + c * x^2
  end
\end{lstlisting}

\begin{figure}[ht]
 	\includegraphics[width =1.\linewidth]{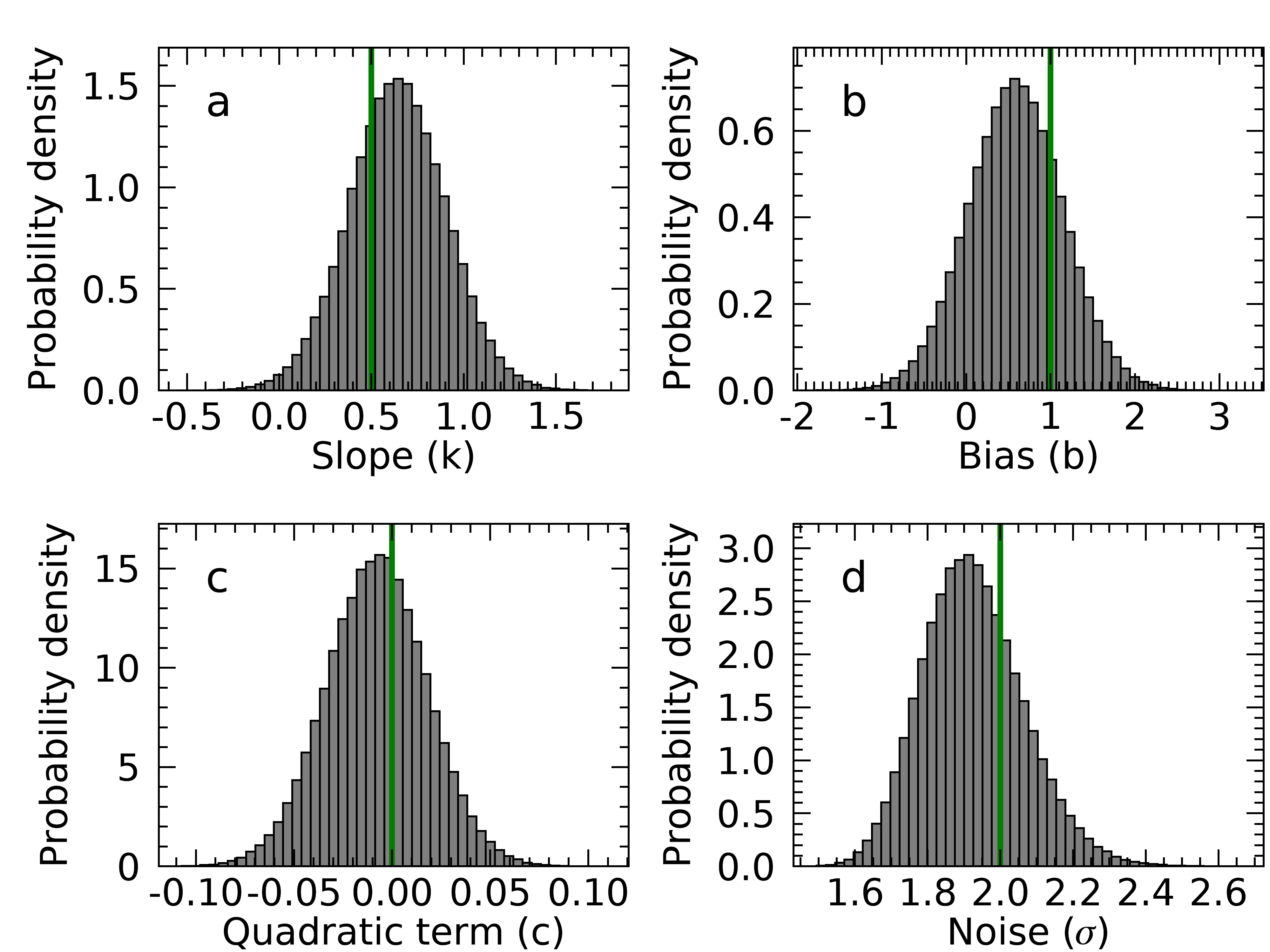}
 	\caption{\textbf{Panel a:} Normalised  histograms approximating marginalised posterior distributions of  the  slope $k$ (a), bias $b$ (b), quadratic term $c$ (c) and noise level $\sigma$ (d) obtained from $10^5$ MCMC samples using the quadratic model $y = k x + b + cx^2$. True values of the parameters  used to generate synthetic data points are shown by vertical green lines.}
 	\label{fig:hist_quad}
 \end{figure}

\begin{figure}[ht]
 \centering
 	\includegraphics[width =0.7\linewidth]{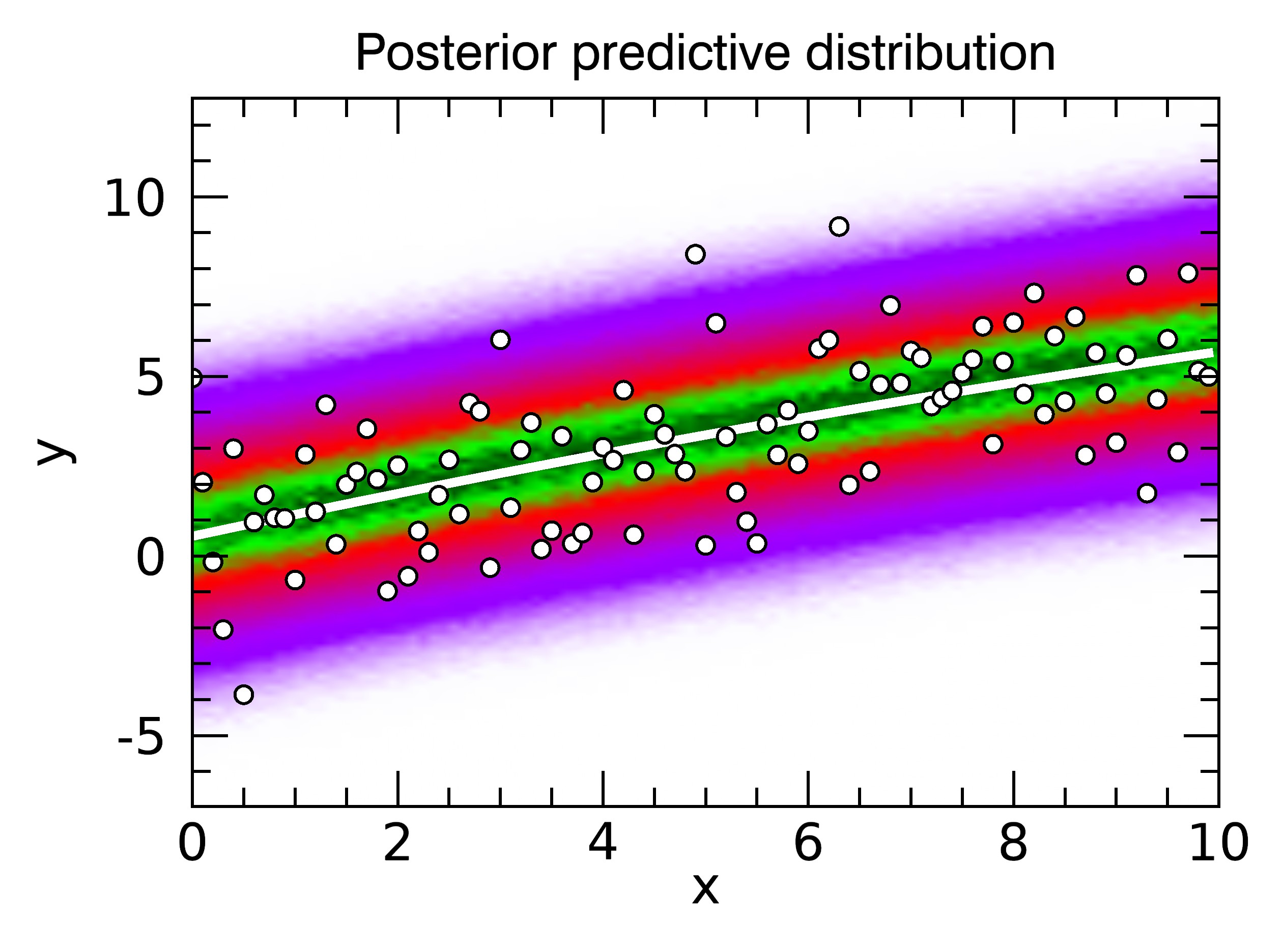}
 	\caption{Posterior predictive probability distribution for a quadratic dependence fitted to the noisy synthetic data using the \texttt{MCMC\_FIT} function. Data points are indicates by white circles while the white line shows the best fit.}
 	\label{fig:ppd_quad}
 \end{figure}

The MCMC Bayesian inference is done for both models and then the models are compared by calculating the Bayes factor.
Figures~\ref{fig:hist_quad} and \ref{fig:ppd_quad} show the MCMC inference results for the quadratic model given by Eq.~(\ref{eq:quad_model}).
Though the best fits and posterior predictive distributions (see Figs~\ref{fig:ppd_linear} and \ref{fig:ppd_quad}) are very similar, the histograms of marginal posterior distributions are found to be significantly broader in comparison with the linear case.
This demonstrates that the additional quadratic term does not improve the fit. The $\chi^2$ and reduced $\chi^2$ metrics are almost the same for both models  (see Table \ref{tbl:model_comparison}) and do not show any significant advantage of one model against the other.

SoBAT includes the \texttt{MCMC\_EVIDENCE} function which allows us to calculate Bayesian evidences and hence the Bayes factor for comparing the models as described in Listing \ref{lst:bayes_factor}, where \texttt{samples\_l} and \texttt{samples\_q} are the MCMC samples simulated using the linear and quadratic models, respectively.
The computed Bayes factor ($28.8$) indicates strong evidence in the favour of the linear model.
This result is expected since we generated the synthetic data using the linear dependence with the background normally distributed noise.

\begin{lstlisting}[float,caption=Calculating the Bayes factor,label={lst:bayes_factor}]
  e_l = mcmc_fit_evidence(samples_l,$
          x, y, priors, 'lin_model')
  e_q= mcmc_fit_evidence(samples_q,$
         x, y, priors, 'quad_model')
  Bayes_factor  = e_l/e_q
\end{lstlisting}

\begin{table*}[]
\centering
\caption{Quantitative comparison of the linear and quadratic models}
\label{tbl:model_comparison}
\begin{tabular}{lcccc}
\hline
\hline
Model & Chi-squared & Reduced chi-squared &  Evidence & Bayes factor\\
\hline
$M_1$: Linear &346.8 &3.539 & $4.7\times10^{-94}$ & $28.8$\\
$M_2$: Quadratic &346.2 &3.569& $1.6\times10^{-95}$ & $-28.8$\\
\hline
\end{tabular}
\end{table*}

\section{Application to realistic problems}
\label{sect:applications}

In this section we illustrate the application of SoBAT to problems in solar physics.

\subsection{Coronal loop seismology using damped kink oscillations}

Coronal loops are frequently observed to perform large amplitude, rapidly-damped, transverse oscillations when perturbed by events such as flares and coronal mass ejections.
Their rapid damping is explained by resonant absorption which causes a transfer of energy from the kink mode to the torsional Alfv\'en mode \citep[e.g. see the recent review by][]{0741-3335-58-1-014001}.
\citet{2013A&A...551A..40P} proposed a method to infer the transverse density profile in the oscillating coronal loop using the shape of the damping profile of the kink oscillation \citep{2013A&A...551A..39H,2012A&A...539A..37P,2015A&A...578A..99P,2016A&A...585L...6P,10.3389/fspas.2019.00022}.
The method was first applied in \citet{2016A&A...589A.136P} using a Levenberg-Marquardt least-squares fit to the data using the \textsc{IDL} code \textsc{MPFIT} \citep{2009ASPC..411..251M}.
It was extended in \citet{2017A&A...600A..78P} to include additional physical effects and also use Bayesian inference.
\citet{2017A&A...607A...8P} also included the presence of a large initial displacement of the loop equilibrium position.
A benefit of the MCMC approach is that we can readily extend our models in this way, allowing us to investigate further details in the data.

We note that in previous applications of our MCMC code to coronal seismology \citep{2017A&A...600A..78P,2017A&A...600L...7P,2017A&A...607A...8P,2017A&A...605A..65G}, posterior summaries were given using the median value (and uncertainties by the 95\% credible interval).
Here, as well as in \citet{2018ApJ...860...31P}, the maximum a posteriori probability (MAP) estimate is used rather than the median.

\begin{figure}[ht]
 	\includegraphics[width =1.\linewidth]{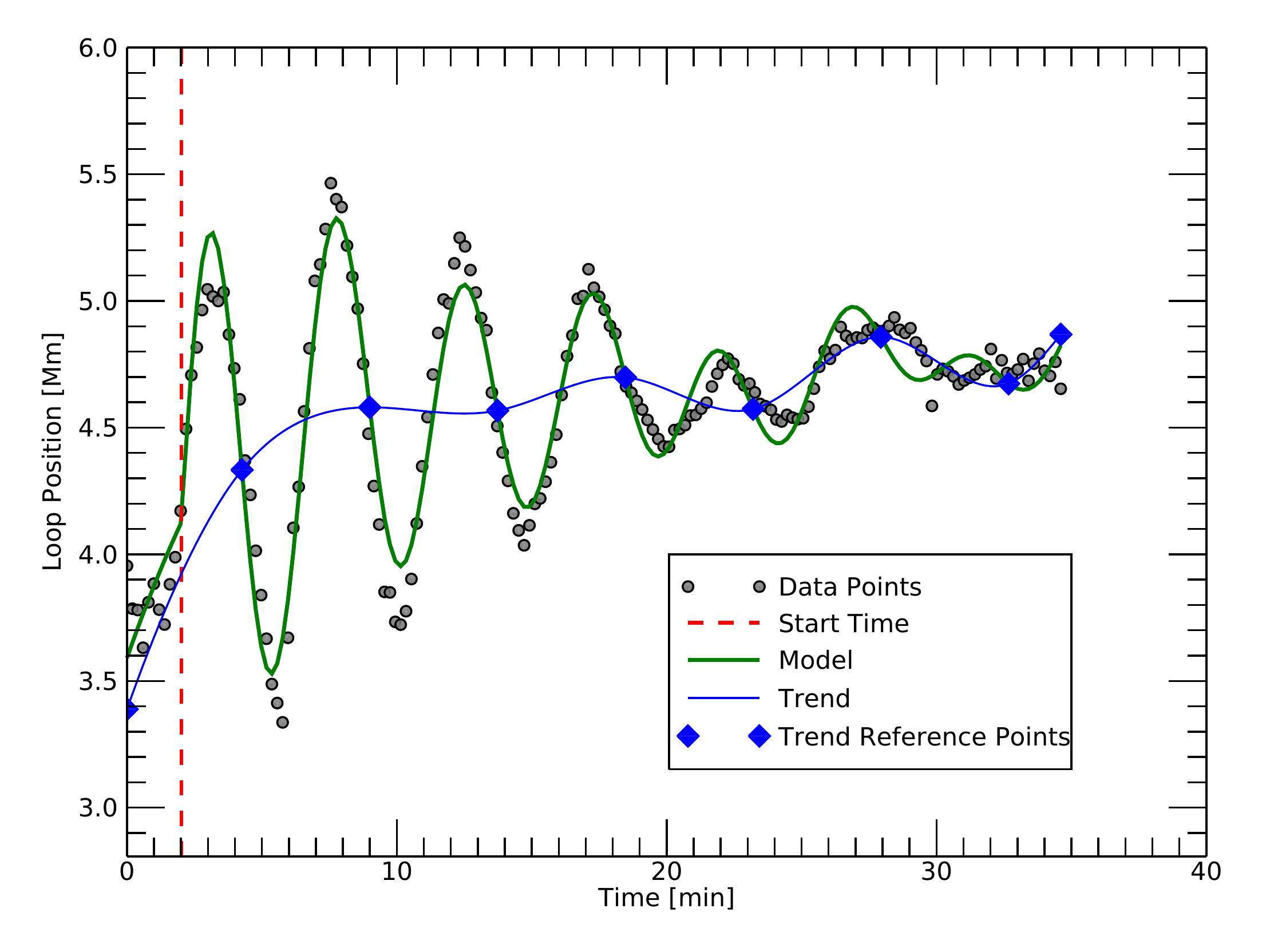}
 	\caption{Best fit (green line) computed for the simplified model of decaying kink oscillations. Observational data points are shown by grey circles. The inferred background trend computed by spline interpolation from the reference points (blue diamonds) is shown by a blue line. The vertical red dashed line denotes the oscillation start time.}
 	\label{fig:exp_osc_fit}
 \end{figure}

In this paper, we use the simplified version of the oscillation profile model published in \citet{2017A&A...600A..78P}:

\begin{equation}
y(t) = y_{\rm tr}(t) + \left\{  \begin{matrix}
A_0 e^{-\left(\frac{\tilde{t}}{\tau}\right)^n} \sin \left( \frac{2\pi \tilde{t}}{P}+ \phi_0 \right) ,& \tilde{t} \ge 0 \\
x_0 ,& \tilde{t} < 0
\end{matrix}\right.,
\end{equation}
where $\phi_0 = \arcsin(\frac{x_0}{A_0})$ is the initial phase, $A_0$ is the initial amplitude, $t_0$ is the  start time of the oscillation,
$\tilde{t}=t-t_0$, 
$P$ is the oscillation period, and $x_0$ is the initial displacement which prescribes the oscillation phase. The parameter $n$ prescribes the damping profile.
The background trend ($y_{\rm tr}$) prescribes the equilibrium position and is calculated using spline interpolation from the reference points located at the time instances when the loop comes through the equilibrium (blue diamonds in Figure \ref{fig:exp_osc_fit}). The positions of the reference points are free parameters of the model and are identified during the Bayesian inference. \textbf{[give listings in appendix]}

\begin{figure}[ht]
 	\includegraphics[width =1.\linewidth]{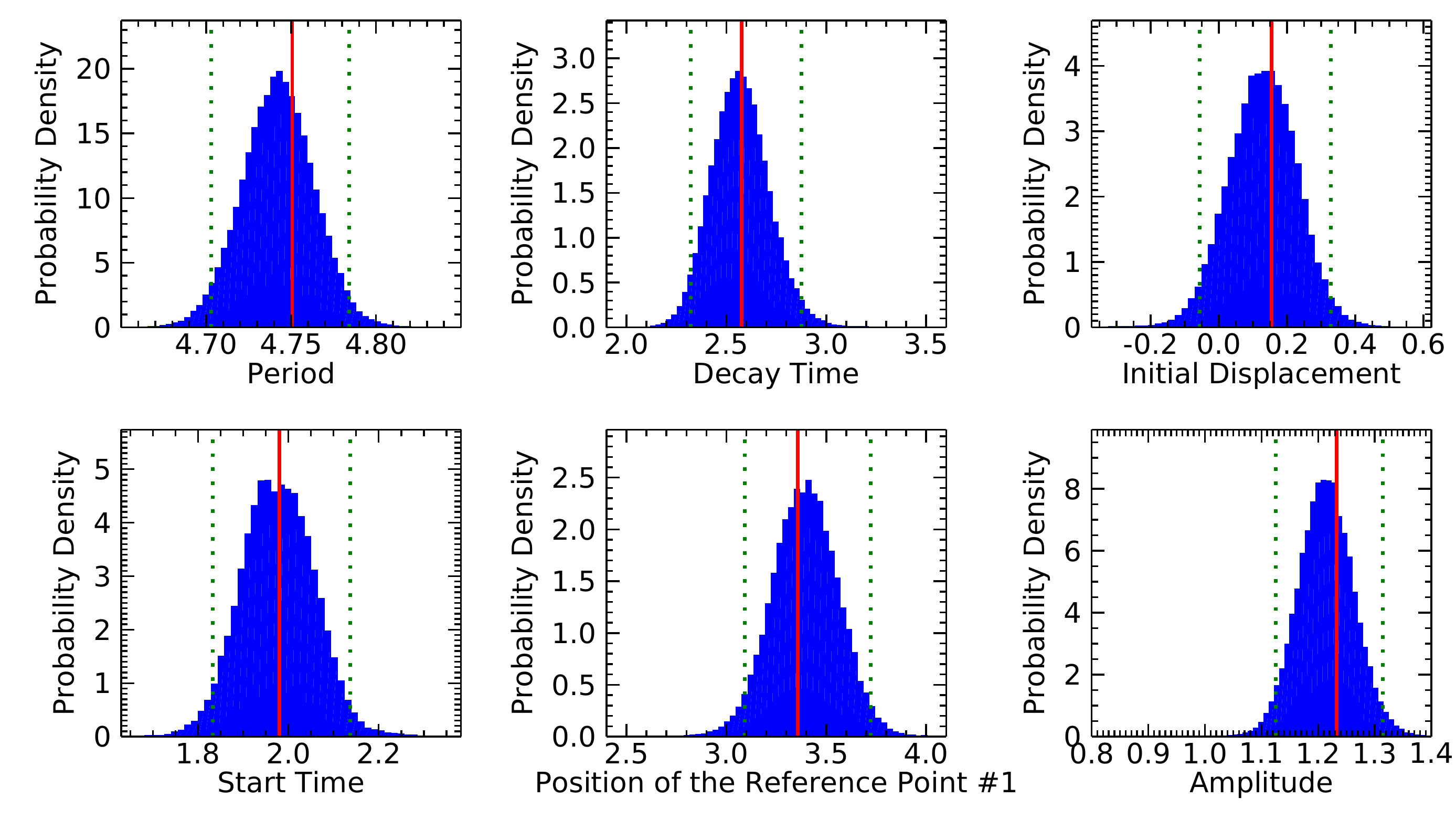}
 	\caption{Histograms approximating marginalised posterior PDF  obtained using the \texttt{MCMC\_FIT} routine \textbf{for the simplified model of exponentially decaying kink oscillations}. The MAP estimates are indicated with the vertical red lines, while the dotted lines show 95\% credible intervals.}
 	\label{fig:exp_osc_hist}
 \end{figure}

As an example, we consider the time series of the loop position taken for Event 43 Loop 3 from the catalogue of oscillations by \citet{2016A&A...585A.137G}.
This loop is also referred to as Loop~\#1 in the seismological analysis by \citet{2016A&A...589A.136P,2017A&A...600A..78P}.
The observational data points and the best fit obtained using the \texttt{MCMC\_FIT} function are shown in Figure \ref{fig:exp_osc_fit}. The histograms approximating marginal posterior distributions of oscillation period, amplitude, decay time, initial displacement, start time, and the position of a trend reference point are given in Figure~\ref{fig:exp_osc_hist}.

\begin{figure}[ht]
 	\includegraphics[width =1.\linewidth]{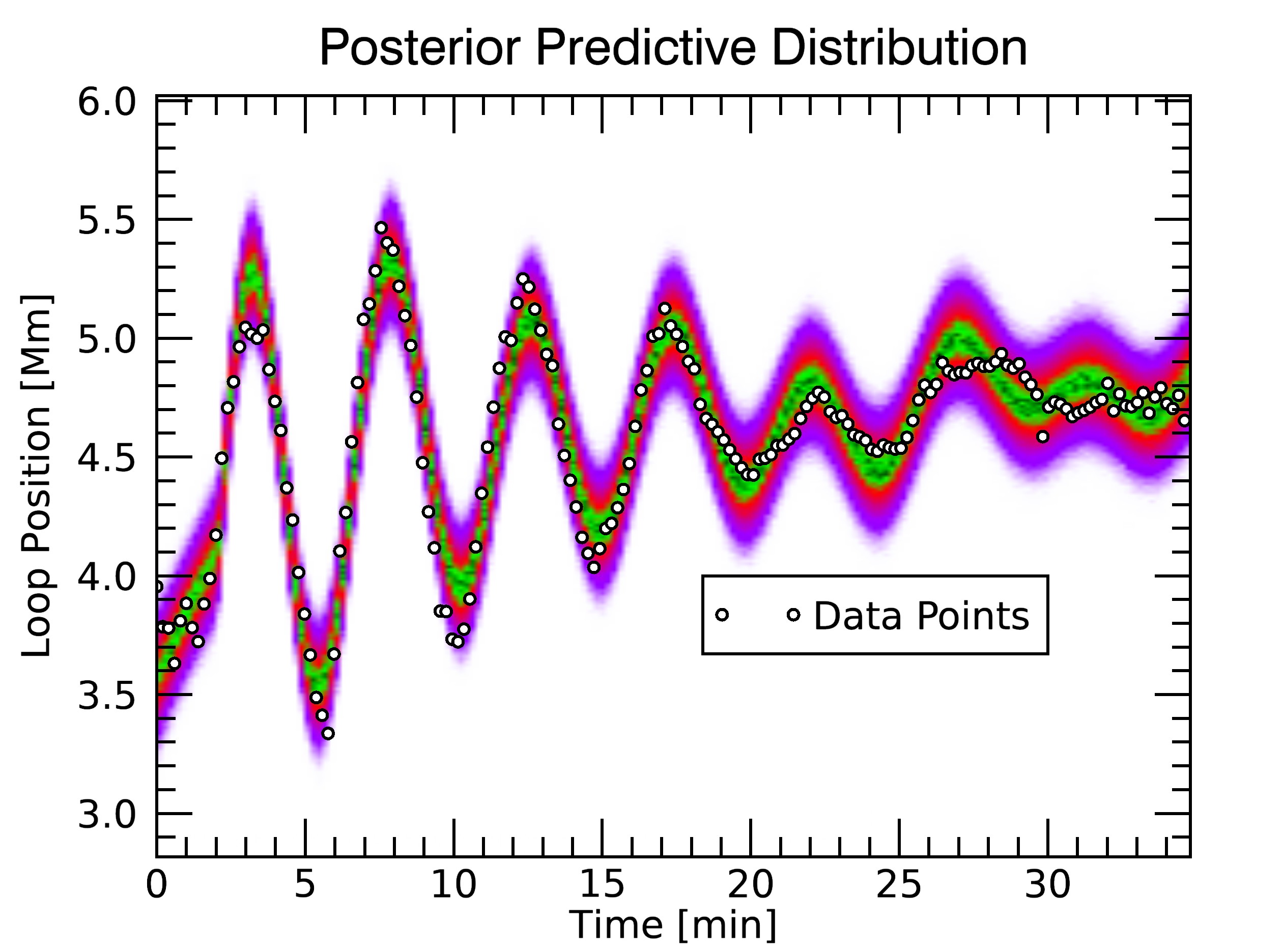}
 	\caption{Posterior predictive distribution PDF (background colour) over-plotted with the observed data points (circles).}
 	\label{fig:exp_osc_ppd}
 \end{figure}

The Posterior Predictive distribution inferred using our MCMC code is given in Figure \ref{fig:exp_osc_ppd}.
The shaded area demonstrates the region on the plot where the data points are predicted to be observed.
For a data consistent inversion, the measured data points should be located inside the shaded region and the shaded area itself should not broaden far away from the data points.
That means that a model should predict the observed data points, but it should not predict observations being far away from the actually observed data.
 
\section{Conclusions}
\label{sect:conclusions}

In this paper, we have described a new code written in IDL to perform MCMC sampling and Bayesian inference for the purpose of testing data against one or more models.
This method and code is applicable to a wide range of problems. It requires that the user supplies a function which returns the predicted values of the data using model parameters, and the prior ranges for these parameters.
These priors may either be prescribed limits for the parameter, or else reasonable estimates for the data being considered.

Since the method is based on forward modelling of the data and efficient sampling of the parameter space it is able to describe model parameters which have arbitrary posterior probability distributions.
This allows reliable estimations of the values and uncertainties of model parameters.
Furthermore, it allows the method to accommodate both well-posed and ill-posed problems.
This is convenient for attempts to reliably extract the maximum information from the available data.
For example, the seismological method of determining the density profile of coronal loops using damped kink oscillations uses the shape of the damping profile to make the problem well-posed.
In the case of the data not supporting a reliable determination of the shape, the problem reverts to being ill-posed and the MCMC sampling recovers an inverse relationship between the density contrast and inhomogeneous layer width \citep[see][for further discussion]{2018ApJ...860...31P}.

Our code has also been used to estimate the density profile of a coronal loop \citep{2017A&A...600L...7P,2018ApJ...860...31P,2017A&A...605A..65G} using a simple procedure for forward modelling the extreme ultraviolet (EUV) emission based on the isothermal approximation \citep[e.g.][]{2007ApJ...656..577A},
and recently applied to the problem of analysing quasi-periodic pulsations in solar and stellar flares \citep{Broomhall_2019}.

The Bayesian evidence may be used to compare two or more competing models for the same data.
In comparison to other tests such as the (reduced) chi-squared, its robustness is increased by considering all prior and posterior information rather than simply the goodness of the model best fits.

The code is available at GitHub page \url{https://github.com/Sergey-Anfinogentov/SoBAT}. According to our knowledge it is the only avialable MCMC code written in IDL which is ready to use out of the box. Example of the code usage in appendix and also available at GitHub.

\acknowledgments

S.A.A. acknowledges the  support of the Russian Scientific Foundation under research grant No 18-72-00144 (the development of the code, sections \ref{sect:method}-\ref{sect:examples}); V.M.N. was supported by the RFBR project No 18-29-21016 (Introduction and Conclusions).
D.J.P. was supported by the European Research Council (ERC) under the European Union's Horizon 2020 research and innovation programme (grant agreement No 724326).
The data is used courtesy of the SDO/AIA team.

%

\vspace{5mm}
\facilities{SDO/AIA}


\software{Interactive Data Language (IDL)}


\appendix

\section{Listing of kink oscillation parameter inference}
\begin{lstlisting}[caption=Running MCMC fitting of Decaying sinusoid into the observed displacements of the oscillating coronal loop,label={lst:kink_mcmc}]
    pro kink_example_data, x, y
    ;observational data points
    x = [0.00,0.20,0.40,0.60,0.80,0.99,1.19,1.39,1.59,1.79,1.99,2.19,$
        2.39,2.59,2.78,2.98,3.18,3.38,3.58,3.78,3.98,4.18,4.38,4.57,$
        4.77,4.97,5.17,5.37,5.57,5.77,5.97,6.16,6.36,6.56,6.76,6.96,$
        7.16,7.36,7.56,7.76,7.95,8.15,8.35,8.55,8.75,8.95,9.15,9.35,$
        9.55,9.74,9.94,10.14,10.34,10.54,10.74,10.94,11.14,11.34,11.53,$
        11.73,11.93,12.13,12.33,12.53,12.73,12.93,13.13,13.32,13.52,$
        13.72,13.92,14.12,14.32,14.52,14.72,14.92,15.11,15.31,15.51,$
        15.71,15.91,16.11,16.31,16.51,16.70,16.90,17.10,17.30,17.50,$
        17.70,17.90,18.10,18.30,18.49,18.69,18.89,19.09,19.29,19.49,$
        19.69,19.89,20.09,20.28,20.48,20.68,20.88,21.08,21.28,21.48,$
        21.68,21.88,22.07,22.27,22.47,22.67,22.87,23.07,23.27,23.47,$
        23.67,23.86,24.06,24.26,24.46,24.66,24.86,25.06,25.26,25.45,$
        25.65,25.85,26.05,26.25,26.45,26.65,26.85,27.05,27.24,27.44,$
        27.64,27.84,28.04,28.24,28.44,28.64,28.84,29.03,29.23,29.43,$
        29.63,29.83,30.03,30.23,30.43,30.63,30.82,31.02,31.22,31.42,$
        31.62,31.82,32.02,32.22,32.42,32.61,32.81,33.01,33.21,33.41,$
        33.61,33.81,34.01,34.21,34.40,34.60]
    y = [3.95,3.79,3.78,3.63,3.81,3.88,3.78,3.72,3.88,3.99,4.17,4.49,$
        4.71,4.82,4.96,5.05,5.02,5.00,5.03,4.87,4.73,4.61,4.37,4.23,$
        4.01,3.84,3.67,3.49,3.41,3.34,3.67,4.10,4.27,4.56,4.81,5.08,$
        5.14,5.28,5.46,5.40,5.37,5.22,5.10,4.97,4.75,4.48,4.27,4.12,$
        3.85,3.85,3.73,3.72,3.78,3.90,4.12,4.35,4.54,4.71,4.87,5.01,$
        4.99,5.15,5.25,5.22,5.12,5.03,4.93,4.88,4.64,4.51,4.40,4.29,$
        4.16,4.09,4.04,4.11,4.20,4.22,4.29,4.36,4.47,4.63,4.78,4.86,$
        5.01,5.02,5.13,5.05,5.02,4.97,4.90,4.87,4.72,4.66,4.64,4.61,$
        4.57,4.53,4.49,4.46,4.43,4.42,4.49,4.50,4.51,4.55,4.55,4.57,$
        4.60,4.66,4.71,4.75,4.77,4.75,4.69,4.67,4.67,4.64,4.59,4.58,$
        4.57,4.53,4.52,4.55,4.54,4.53,4.54,4.58,4.65,4.74,4.80,4.77,$
        4.81,4.90,4.86,4.85,4.86,4.85,4.88,4.89,4.88,4.88,4.90,4.94,$
        4.89,4.87,4.89,4.84,4.80,4.76,4.59,4.71,4.73,4.72,4.70,4.67,$
        4.69,4.70,4.71,4.73,4.74,4.81,4.69,4.77,4.72,4.71,4.73,4.77,$
        4.68,4.75,4.79,4.72,4.70,4.76,4.65]
    end

    ; model function accepts keyword parameter N_TREND
    ; that will be passed to it
    function model_exp_decay, x, a, n_trend=n_trend
      tstart = a[0]             ; oscillation start time
      period = a[1]             ; oscillation period
      q_factor = a[2]           ; oscillation decay_time
      amp = a[3]                ; initial amplitude
      displ = a[4]
      ref_y = a[5:5+n_trend-1]  ; trend reference points
      ref_x = linspace(x[0],x[-1],n_trend)  
      tau = q_factor*period     ; decay time
      tosc = x-tstart
      omega = 2.d*!dpi/period
      phi = asin((displ))       ; initial phase
      ; decaying profile
      damp = amp * exp(-(tosc/tau)^1) * (x ge tstart)
      oscillation = damp * sin(omega*(tosc>0d) + phi)
      trend = spline(ref_x, ref_y, x)
      return, trend + oscillation
    end

    pro example_kink
      kink_example_data, x, y
      plot, x, y, /psym
      ; use 5 reference points for the trend
      n_trend = 5
      ; initial values
      pars = [1d, 2d, 2d, 1d, 0d, 5d, 5d, 5d, 5d, 5d]
      priors = [$
        prior_uniform(0d, 5d), $   ; start time
        prior_uniform(1d, 10d), $  ; period
        prior_uniform(1d, 10d), $  ; q factor
        prior_uniform(0d, 10d), $  ; amplitude
        prior_uniform(-1d, 1d), $  ; initial displacement
        ; trend reference points
        replicate(prior_uniform(min(y), max(y)), n_trend) $
      ]
      model = 'model_exp_decay'
      ; sample posterior distribution using the MCMC
      y_fit = mcmc_fit(x, y, pars, model, n_samples = 100000l, priors=priors, $
        burn_in=50000l, samples=samples, n_trend=n_trend)
    end
\end{lstlisting}




\bibliographystyle{aasjournal}



\end{document}